\documentclass[superscriptaddress,onecolumn]{revtex4}

\usepackage{caption}
\usepackage{subcaption}
\usepackage{amssymb}
\usepackage{amsfonts}
\usepackage{amsmath}
\usepackage{graphicx}
\usepackage{dcolumn}
\usepackage{bm}
\begin{document}
\title{Local symmetries in $f(T)$-like models: lessons from 2D}

\author{Franco Fiorini}
\email{francof@cab.cnea.gov.ar} \affiliation{Departamento de Ingenier\'{i}a en Telecomunicaciones, Consejo Nacional de Investigaciones Cient\'{i}ficas y T\'{e}cnicas (CONICET) and Instituto Balseiro (UNCUYO), Centro At\'{o}mico Bariloche, Av. Ezequiel Bustillo 9500, CP8400, S. C. de Bariloche, Rio Negro, Argentina.}


\begin{abstract}
The comprehension of the intricate structure associated to the local symmetries encoded in the tetrad field, as well as its physical meaning, is perhaps the most important unsolved problem within $f(T)$ gravity. This is inextricably connected to the number, nature and potential impact that the additional degree/s of freedom might have within these --and other closely related--models of gravity in which the local Lorentz invariance is broken at some level. Here we review and further explain some recent results which make use of the more placid scenery provided by 2D-torsional models of gravity, where the local symmetries adapted to a given geometry can be fully characterized.
\end{abstract}
\maketitle

\section{Introduction}\label{sec1}
This work can be thought as a tentative to make some sense out of the internal symmetries lying behind $f(T)$ gravity \cite{FF}-\cite{Review}. For this endeavor to be fruitful, it is crucial to conceive $f(T)$ gravity as a gravitational theory dictating the dynamics of the entire tetrad (vierbein) field $e^a$ (\emph{pure tetrad} approach), instead of thinking of gravity in purely \emph{metric} terms. This means that one has to put the effort to understand the additional gravitational degrees of freedom from a wider perspective, not involving the metric tensor $\bold{g}=\eta_{ab}e^ae^b$ alone, but the tetrad field giving rise to that metric too. Of course, this practice ultimately entails the abandonment of the full local Lorentz group as a symmetry group, at least in the regime where $f(T)$ gravity notedly differs from General Relativity (GR) written in its teleparallel equivalent form. Originally, $f(T)$ gravity was discussed in the context of high-energy (strong-field) deformations of GR of the sort $f(T)=T+T^2/\lambda+O(T^3/\lambda^2)$. Here, $\ell=\lambda^{-1/2}$ is the length scale at which the Lorentz symmetry is broken: for instance, in Refs. \cite{FF} and \cite{FF2} it was found that $H_{max}^2\propto\lambda$ is the maximum Hubble rate characterizing the early inflationary era of regular, FRW-like cosmological models. The scale $\ell$ also appears in the context of solutions describing vacuum, regular black holes interiors, see. \cite{BF1} and \cite{BF2}.

The proposed point of view (at least, conceptually) is appreciably different from the one adopted in the so called \emph{covariant} formulation \cite{covfor}, see also \cite{covfor2}. The latter intends to recover the Lorentz covariance by introducing a non-dynamical connection, therefore, in a way, it leads to a more metric-driven kind of theory. However, it was shown recently that the covariant formulation is not \emph{that} covariant after all \cite{grupo23}, thereby for better or worse one has to accept the fact that the breaking of the Lorentz symmetry is a characteristic feature of $f(T)$ gravity, at least in the strong field regime. Having said that, it is true that we cannot simply downplay the role of the metric tensor in what concerns to the description of the gravitational field, so we are in a somewhat awkward position in which the structure of the metric serves us just as a guiding principle with the purpose of unraveling the more primary layout present in the tetrad.

In order to delve on this facts, we propose to rise --and hopefully answer-- the two following unanswered questions:

\bigskip

(a) If we conceive a certain spacetime locally described in the standard (\emph{\`{a} la} GR) terms by a metric $\bold{g}$, what is the full set $\mathcal{E}(\textbf{g})$ of tetrads representing that local geometry? In other words, what is the full set of tetrads solving the $f(T)$ motion equations adapted to a given local geometry encoded in $\bold{g}$?

(b) In case that we convince ourselves that we have an answer for question (a), what physical meaning (if any) can we bestow to the set $\mathcal{E}(\textbf{g})$? Or, in simpler terms, is there any more than \emph{geometry} contained in $\mathcal{E}(\textbf{g})$?

\bigskip
A partial answer to question (a) is already available through the concept of \emph{remnant symmetries}. These are defined as the local Lorentz transformations (LLT) leaving invariant the Weitzenb\"{o}ck pseudo-scalar $T$ \cite{grupo}. We recall that $T$ is a scalar only under general coordinate changes and global Lorentz transformations of the tetrad field. Nevertheless, after finding a certain $e^b$ obeying the motion equations and constructing its corresponding $T(e^b)$, it is possible in principle to find LLT $\Lambda^a_{b}(x^{\mu})$ in such a way that the whole class of tetrads $e^a=\Lambda^a_{\,\,b}(x^{\mu})\, e^b$ verify $T(e^a)=T(e^b)$. These constitute remnant local symmetries leaving invariant the $f(T)$ action; we shall call this set $\mathcal{A}(e^a)$. As mentioned, this set provides just a \emph{partial} answer to question (a), because the same local geometry $\textbf{g}$ might be described by tetrads leading to different values of $T$, and then, by definition, not connected by the action of the remnant symmetries. But we have to start somewhere. To read more about the remnant symmetries, we suggest also the articles \cite{Nos5}-\cite{Manu}.

Needless to say, question (b) is more subtle, and its potential answer, more evasive, because it depends on the just partially answered question (a). Sensible answers to this question should give us important information about the additional degrees of freedom arising in these non-local Lorentz invariant, torsional theories of gravity.

In this work, which constitutes an expanded version of the talk given at Tartu, we will develop some strategy for dealing with both questions raised above, and for understanding, to certain extent, the relation between $\mathcal{A}(e^a)$ and $\mathcal{E}(\textbf{g})$. In order to tackle this problem, we will construct several two-dimensional, non-local Lorentz invariant models relaying on absolute parallelism, and then analyze the local symmetries arising in --what we think are-- situations not dearth of physical interest. Hopefully the results and insights coming from these 2D toy models could serve to trigger other studies following the same lines on realistic, four-dimensional circumstances. The oncoming exposition is a continuation of the recent article \cite{Andronikos}.

\section{Motivation: Milne, Rindler, and near horizon geometry}\label{sec2}
Two dimensional gravity has, undoubtedly, the advantage of simplicity. One can perform many exact computations and push forward ideas which will be supported by exact results. However, it is a fact that most of lower dimensional results are not directly applicable to real, physical situations; moreover, it is well known that many of those results are simply wrong in higher dimensions, so it is clever not to put too much faith in those docile and tractable outcomes of lower dimensional gravity. Nonetheless, we shall review in this section why 2D Milne and Rindler spaces are instrumental in describing the near-horizon Schwarzschild geometry experienced by a radial free-falling observer. Throughout this section, we will deliberately abide to a pure metric point of view, without making any mention to the underlying tetrad field; hopefully this will allow us to to transit the trek from the metric-based approach, to the more enlightening one based on the structure of the entire tetrad field, to be treated in Section \ref{sec4}

From an interior point of view (this is, for $r<2M$ in standard, Schwarzschild coordinates), Schwarzschild spacetime looks quite alike a cosmological model; in fact, Kantowski and Sachs \cite{KS66} showed that the metric
\begin{align}
  ds^2 = -dt^2+a^2(t)\,dx^2+b^2(t)\,d\Omega^2,
  \label{metrica}
\end{align}
represents the interior Schwarzschild spacetime, provided
\begin{align}
  a(t)=a_{1}\,\tan\left[N(t)\right],\quad
  b(t) = b_{1}\,\cos\negmedspace^2\left[N(t)\right].
  \label{func-a}
\end{align}
Here the two scale factors are written in terms of the time function $N(t)$ defined implicitly by
\begin{align}
 t-t_{0} =b_{1} (N+\sin N\cos N),
  \label{imptime}
\end{align}
 and the integration constants $a_{1}$ and $b_{1}$ are constrained according to $-\pi/2\leq a_{1}<0$ and $b_{1} \neq 0$. The black hole horizon corresponds to the limit $N\rightarrow 0$, for which the function $a(t)$ goes to zero. In this way, after fixing $t_{0}=0$, near the horizon we can invert (\ref{imptime}) and obtain $N(t)\approx t/2 b_{1}$, while the scale factors at the lowest order result
\begin{align}
  a(t)\approx a_{1}\,t ,\quad
  b(t) \approx b_{1}.
  \label{scalelow}
\end{align}
The geodesic motion experienced by a radial observer near the horizon might then be studied by using the 2D line element
\begin{equation}\label{metint}
ds^2=-dt^2+a_{1}^2\,t^2dx^2.
\end{equation}
The metric (\ref{metint}) can be viewed as a two-dimensional version of Milne metric. Let us remember that Milne space is a special case of $K=-1$, FRW spaces in which the scale factor is linear in time. Only in this case the resulting space is actually a portion of Minkowski space in disguise. A radial free falling observer near the horizon (from the interior point of view), will experience thus a locally flat space which differs from Minkowski space only in its global properties. Precisely, after changing to the conformal time $\eta(t)=\pm\, a_{1}^{-1}\log(t)$, the metric (\ref{metint}) reads $ds^2=a_{1}^{2}\exp{(2 a_{1} \eta)}[-d\eta^2+dx^2]$, which in turn can be converted into $ds^2=-(dx^{0})^2+(dx^{1})^2$ once we have defined
\begin{equation}
  x^{0}= \exp{(a_{1} \eta)}\cosh(a_{1} x),\,\,\,\,   x^{1}= \exp{(a_{1} \eta)}\sinh(a_{1} x).
 \label{coormil}
\end{equation}
Notice that $0<x^{0}<\infty$, $-\infty<x^{1}<\infty$, and $x^{0}>x^{1}$. Of course, the $4$-metric is curve even in the vicinity of the horizon, and reads $ds^2=-(dx^{0})^2+(dx^{1})^2+b_{1}^2d\Omega^2$.

There is another point of view in what concerns to the description of the near horizon geometry, i.e., the one that would adopt an observer experiencing the situation from the outside. In this case the local geometry is described by a Rindler-like metric. Actually, near the horizon the Schwarzschild metric looks in appropriate coordinates
\begin{equation}
ds^2=- (1-X/4M)^2\,dT^2+dX^2+4M^2d\Omega^2,
\label{rindler}
\end{equation}
where $M$ is the black hole mass. It is well know that the $T,X$ sector of metric (\ref{rindler}) corresponds to the \emph{Rindler wedge}, that is to say, the portion of Minkowski space $ds^2=-(dx^{0})^2+(dx^{1})^2$ verifying $x^{1}>|x^{0}|$.

For purposes to be clarified later, it is important to see how 2D Rindler metric can be obtained from the other side of the horizon by starting from the 2D-Milne line element (\ref{metint}), i.e., to establish the equivalence
\begin{equation}
ds^2=-dt^2+a_{1}^2\,t^2dx^2\,\,  \underbrace{\Longrightarrow}_{crossing\, the\, horizon} \,\,  ds^2=- (1-X/4M)^2\,dT^2+dX^2.
\label{rindler2}
\end{equation}
We can do this in three stages by performing simple coordinates changes:

\bigskip

\emph{Stage one}: from $(t,x)$ to $(\bar{T},x)$ (converting Milne into interior-Schwarzschild). Let us take
\begin{equation}
   a_{1}t=\sqrt{2a_{1}\bar{T}-1},\,\,\Longrightarrow\,\,  ds^2=-\frac{d\bar{T}^2}{(2a_{1}\bar{T}-1)}+(2a_{1}\bar{T}-1)\,dx^2,
\label{change1}
\end{equation}
where, clearly, $\bar{T}>1/2a_{1}$.

\emph{Stage two}: from $(\bar{T},x)$ to $(T,\bar{x})$ (crossing the horizon). Here we have
\begin{equation}
 T=x,\,\,\,\bar{x}=\bar{T}   \,\,\Longrightarrow \,\,    ds^2=- (1-2a_{1}\bar{x})\,dT^2+\frac{d\bar{x}^2}{(1-2a_{1}\bar{x})}\,.
\label{metboostchange2}
\end{equation}

\emph{Stage three}: from $(T,\bar{x})$ to $(T,X)$  (converting the exterior-Schwarzschild into Rindler). Finally,
\begin{equation}
   1-a_{1}X=\sqrt{1-2a_{1}\bar{x}}, \,\,\Longrightarrow \,\, ds^2=- (1-a_{1}X)^2\,dT^2+dX^2.
\label{change3}
\end{equation}
Comparing the final form of the metric in (\ref{change3}) with the $T,X$ sector of (\ref{rindler}), we see that $a_{1}=1/4M$, then $a_{1}$ is exactly the surface gravity of the Schwarzschild black hole. The correspondence (\ref{rindler2}), simply as it is, teach us an important lesson: 2D cosmological models having Milne asymptotics, are adequate to describe the near-horizon geometry experienced by radial observers, not only from the interior point of view, but from the exterior as well. This observation motivates the construction of two-dimensional, non local Lorentz invariant gravitational models based on the \emph{diad} (zweibein) field $e^a$, and the study of their potential cosmological solutions having Milne asymptotics. Hopefully, the characterization of such solutions will permit us to better understand the sets $\mathcal{E}(\textbf{g})$ and $\mathcal{A}(e^a)$, where $\textbf{g}$ represents Milne space, and to say something solid in what concerns to question (b) raised in Sec. \ref{sec1}.

\section{2D gravity based purely on torsion}

It is not hard to create 2D models of gravity based on the torsion, and to subsequently force the breaking of the Lorentz invariance by introducing a sort of \emph{Weitzenb\"{o}ck gauge} in which the spin connection is zeroed, as needed in the pure diad approach. In this way, the torsion is just $T^{a}=de^a$ and its local components read $T^{a}_{\,\,\mu\nu}=\partial_{\nu}e^a_{\,\,\mu}-\partial_{\mu}e^a_{\,\,\nu}$. It is important to remember that the Riemann-Weitzenb\"{o}ck link, $T=-R+2e^{-1}\partial_{\nu}(eT_{\mu}^{\,\,\,\mu\nu})$, becomes more or less trivial in 2D
\begin{equation}\label{RWlink}
R=2e^{-1}\partial_{\nu}(eT_{\mu}^{\,\,\,\mu\nu}),
\end{equation}
because of the fact that $T=S^{a}T_{a}=0$ due to the null character of the superpotential $S^{a}$ in two spacetime dimensions. The equation (\ref{RWlink}) is the manifestation of the topological nature of $R$ in two dimensions. We will consider thus the object
\begin{equation}\label{escalardew}
\mathbb{T}=T^{a}_{\,\,\mu\nu}T_{a}^{\,\,\mu\nu},
\end{equation}
which is an invariant under coordinate changes and global Lorentz transformations, but not under LLT, just like $T$. Having this building block in mind, the construction of non-local Lorentz invariant gravitational actions in 2D is merely a matter of mimicking some of the 4D models in vogue, as $f(T)$ gravity and Born-Infeld (determinantal) gravity developed in \cite{BID} and \cite{Vatu}. Along the lines of the former, we propose
\begin{equation}\label{accionefe}
I_{f}=\frac{1}{2\kappa } \int\,e \,f(\mathbb{T})\,d^2x\,+I_{matter},
\end{equation}
where $\kappa$ is a coupling constant with units of squared length. In turn, the latter calls for the 2D-Born-Infeld like action \cite{Vatu2}
\begin{equation}\label{ac2d}
I_{BI}=\frac{\lambda}{2\kappa} \int\, \Big[\sqrt{\mid g_{\mu\nu}+2\lambda^{-1}F_{\mu\nu}\mid}-\sqrt{\mid g_{\mu\nu}\mid}\Big]\,d^2x\,+I_{matter},
\end{equation}
where $\mid(...)\mid$ is a shorthand for the absolute value of the determinant of $(...)$, and $\lambda$ is the BI constant. The tensor $F_{\mu\nu}$ reads
\begin{equation}\label{efe}
F_{\mu\nu}= \alpha\, A_{\mu\nu}+\beta\, B_{\mu\nu}+\gamma\, C_{\mu\nu},
\end{equation}
where in 2D the different contributions are
\begin{equation}\label{pieces}
A_{\mu\nu}=T_{\mu\sigma\rho}T_{\nu}^{\,\,\,\sigma\rho},\,\,\,
B_{\mu\nu}=T_{\sigma\mu\rho}T^{\sigma\,\,\,\rho}_{\,\,\,\nu},\,\,\,
C_{\mu\nu}=g_{\mu\nu} \mathbb{T}.
\end{equation}
These three tensors are designed to verify $Tr(A_{\mu\nu})=Tr(B_{\mu\nu})= \mathbb{T}$, being $C_{\mu\nu}$ a pure trace term, so we have $Tr(F_{\mu\nu})=(\alpha+\beta+2\gamma)\mathbb{T}$.

The equations of motion associated to the action (\ref{accionefe}) are obtained by varying with respect to the diad components $e^{a}_{\mu}$, and they are
\begin{equation}
-4\bigl[T^{\rho\mu a} T_{\rho\mu\nu}-e^{-1}\partial_\mu(e\,T^{a\mu}_{\,\,\,\,\,\,\,\nu})\bigr]
 f^{\prime }-4\,T^{a\mu}_{\,\,\,\,\,\,\,\nu} \partial_\mu \mathbb{T}\, f^{\prime \prime }+e_{\,\,\nu}^a f = 2\kappa\, T_{\,\,\nu}^{a},
 \label{ecuaciones}
\end{equation}
where $f^{\prime}$ and $f^{\prime\prime}$ refers to derivatives with respect to $\mathbb{T}$. In the RHS of (\ref{ecuaciones}), $T_{\,\,\nu}^{a}$ is the diad-projected energy momentum tensor $T_{\,\,\nu}^{a}=e^a_{\,\,\mu}T^{\mu}_{\,\,\nu}$. These equations have an uncanny resemblance with the 4D-$f(T)$ equations; actually, they can be obtained from the $f(T)$ equations by replacing the superpotential $S^{a}_{\,\,\,\mu\nu}$ by the torsion $T^{a}_{\,\,\,\mu\nu}$ itself. We see then that, despite the simplicity of this model, we can augur a non trivial and rich dynamical content.

In what respect to BI gravity, by factoring out $e=\sqrt{\mid g_{\mu\nu}\mid}$ in (\ref{ac2d}), we see that the action integral can be written in the equivalent form
\begin{equation}\label{ac2d2}
I_{BI}=\frac{\lambda}{2\kappa} \int\,e \,\Big[\sqrt{\mid \mathbb{I}+2\lambda^{-1}\mathbb{F}\mid}-1\Big]\,d^2x\,+I_{matter},
\end{equation}
where $\mathbb{I}$ is the identity and $\mathbb{F}=F_{\mu}^{\,\,\nu}$. We can explicitly write down the determinant as
\begin{equation}\label{detee}
\mid \mathbb{I}+2\lambda^{-1}\mathbb{F}\mid=1+2\lambda^{-1}\,Tr(\mathbb{F})+2\lambda^{-2}\,[Tr^2(\mathbb{F})-Tr(\mathbb{F}^2)].
\end{equation}
With the use of expressions (\ref{efe}) and (\ref{pieces}) we can compute the two ingredients appearing in (\ref{detee}), namely
\begin{equation}\label{trazas}
Tr(\mathbb{F})=(\alpha+\beta+2\gamma)\mathbb{T},\,\,\,\,\,\,\,Tr(\mathbb{F}^2)=D_{\mu}^{\,\,\nu}D_{\nu}^{\,\,\mu}+\frac{\gamma}{4}(\gamma+2\alpha+2\beta)\,\mathbb{T}^2,
\end{equation}
where $D_{\mu}^{\,\,\nu}=\alpha A_{\mu}^{\,\,\nu}+\beta B_{\mu}^{\,\,\nu}$ and we used the fact that $A_{\mu}^{\,\,\nu}C_{\nu}^{\,\,\mu}=B_{\mu}^{\,\,\nu}C_{\nu}^{\,\,\mu}=\mathbb{T}^2$ in the derivation of the second equation (\ref{trazas}). In this way, the determinant (\ref{detee}) reads
\begin{equation}\label{deteco}
\mid \mathbb{I}+2\lambda^{-1}\mathbb{F}\mid=1+2\lambda^{-1}(\alpha+\beta+2\gamma) \,\mathbb{T}+2\lambda^{-2}\,\gamma \,\mathbb{T}^2- 2\,\delta\, \lambda^{-2}\,D_{\mu}^{\,\,\nu}D_{\nu}^{\,\,\mu},
\end{equation}
being $\delta=\delta(\alpha,\beta,\gamma)$ a constant which is not important at the moment. We see thus that BI gravity in general is not a sort of masked $f(\mathbb{T})$ theory, due to the explicitly ``non-scalar" contribution of the $D_{\nu}^{\,\,\mu}$ term at order $\lambda^{-2}$. On the other hand, if we normalize according to $\alpha+\beta+2\gamma=1$, at the lowest order we obtain the action
\begin{equation}\label{accionlow}
I_{\downarrow}=\frac{1}{2\kappa} \int\,e \,\mathbb{T}\,d^2x\,+I_{matter},
\end{equation}
which of course, it is also the simplest $f(\mathbb{T})$ theory we can conceive in two spacetime dimensions; the motion equations associated to this action are easily obtained from (\ref{ecuaciones}) by considering $f(\mathbb{T})=\mathbb{T}$, and they are
\begin{equation}
    \mathbb{T}\,e^{a}_{\, \,\nu}  -4\, T^{\sigma\rho a}T_{\sigma\rho \nu}  +4\, e^{-1} \partial_{\mu} \left( e\, T^{a\mu}_{\,\,\,\,\,\, \,\nu} \right) = 2\kappa\, \, T^{a}_{\, \,\nu}.
\label{eq: ecTT}
\end{equation}
This would be our basic working model for non-local Lorentz invariant, two-dimensional gravity based on the torsion tensor alone. Curiously enough, this model contains a local Lorentz invariant sector, as can be seen from the equations (\ref{eq: ecTT}). Contracting these equations with the inverse diad $e_{a}^{\, \,\nu}$ we have
\begin{equation}
    -2\,\mathbb{T}+4\, e^{-1} \partial_{\mu} \left( e\, T^{a\mu}_{\,\,\,\,\,\, \,\nu} \right)e_{a}^{\, \,\nu} = 2\kappa\,\bold{T},
\label{eq: ecTTcon}
\end{equation}
where $\bold{T}=T^{a}_{\, \,\nu}\,e_{a}^{\, \,\nu}$. Unfortunately, we found ourselves in the need of introducing yet another $T$-symbol, this time, not related to the torsion $T^a$, but to the energy-momentum tensor $T^{a}_{\mu}$ instead. We hope the context will be enough for distinguishing the various $T$-related objects appearing along these lines. 

 We see that $\bold{T}$ is a scalar under LLT, because $\bold{T}=T_{\mu\nu}e_{b}^{\, \,\mu}e_{a}^{\, \,\nu}\eta^{ab}=T_{\mu\nu}g^{\mu\nu}=Tr(T_{\mu\nu})$. On the other hand, the second term of the LHS can be written as
\begin{eqnarray}
  4\, e^{-1} \partial_{\mu} \left( e\, T^{a\mu}_{\,\,\,\,\,\, \,\nu} \right)e_{a}^{\, \,\nu}&=
  &4\, e^{-1} \partial_{\mu} \left( e\,T_{\nu}^{\,\,\,\mu\nu}\right)-4\,T^{a\mu}_{\,\,\,\,\,\, \,\nu}\,\partial_{\mu}e^{\,\,\nu}_{a}\label{calc1}\\
  &=&-2R-2\,T^{a\mu}_{\,\,\,\,\,\, \,\nu}\, (\partial_{\mu} e_{a}^{\, \,\nu}-\partial_{\nu} e_{a}^{\, \,\mu})  \label{calc2}\\
 &=&2(-R+\mathbb{T})\label{calc3}.
\end{eqnarray}
To arrive at (\ref{calc2}) from (\ref{calc1}), we have used (\ref{RWlink}) and the fact that $T^{a\mu}_{\,\,\,\,\,\, \,\nu}$ is antisymmetric in $\mu$ and $\nu$; this enabled us to antisymmetrize $\partial_{\mu}e^{\,\,\nu}_{a}$ and to get another $T_{a\mu}^{\,\,\,\,\,\, \,\nu}$ to finally arrive at (\ref{calc3}). We see that Eq. (\ref{eq: ecTTcon}) ends up being
\begin{equation}
   R = -\kappa\,\bold{T},
\label{JT}
\end{equation}
which is no other than Jackiw-Teitelboim (JT) model without $\Lambda$-term \cite{Claudio1}, \cite{Roman1} (see also \cite{Claudio} and \cite{Roman}). In this way, JT gravity appears as a local Lorentz invariant sector of the pure-torsion theory (\ref{accionlow}).

\section{Near horizon symmetries}\label{sec4}

\subsection{Comparing the sets $\mathcal{E}(\textbf{g})$ and $\mathcal{A}(e^a)$}

Once we have presented several models for 2D-gravity, it is our intention to characterize the sets $\mathcal{E}(\textbf{g})$ and $\mathcal{A}(e^a)$, where $\textbf{g}$ represents Milne space. We are going to do this for the simplest 2D theory so far treated, this is, the one given by the action (\ref{accionlow}). It is interesting to develop this subject in a staggered way, with the purpose of trying to view things first more alike as in the metric mode, and to build upon that later in order to understand $\mathcal{E}(\textbf{g})$ and $\mathcal{A}(e^a)$.

Milne space is a particular case of a cosmological model with metric
\begin{equation}\label{metcos}
ds^2=-dt^2+a^2(t)dx^2,
\end{equation}
where $a(t)=a_{1}\,t$. Then, realizing that our gravitational action is constructed from the diad and its first derivative, we immediately propose a zweibein with the structure $e^a_{\mu}=diag(1,a(t))$, which certainly conduces us to the line element (\ref{metcos}). It is a simple matter to compute the only non-null component of the torsion tensor, $T^{x}_{\,\,\,\,t\,x}=H$, where $H=\dot{a}/a$ is the 2D Hubble rate, and the corresponding $\mathbb{T}=2H^2$. To obtain the motion equations (\ref{eq: ecTT}) is a bit more laborious, but at the end we get
\begin{equation}
H^2=\kappa \rho,\,\,\,\,\,\, H^2+\dot{H}=\kappa(\rho-p)/2.\label{ecsgr2}
\end{equation}
Here, $\rho$ and $p$ are the energy density and pressure of the perfect fluid acting as a source of the equations ($T^{\mu}_{\nu}=diag(-\rho,p)$). We see that our model is not that unrealistic after all, it reproduces Friedmann equation in a very simple setting. As usual, Eqs. (\ref{ecsgr2}) can be combined to obtain the conservation law
\begin{equation}\label{conservacion}
\dot{\rho}+H(\rho+p)=0.
\end{equation}
Additionally, if we impose $p=\omega\, \rho$, this equation can be integrated to get
\begin{eqnarray}\label{energyden}
\rho(t)=\rho_{0}\,a(t)^{-(1+\omega)},
\end{eqnarray}
where $\rho_{0}$ is a constant. In this way, Friedmann equation and (\ref{energyden}) combine to give
\begin{equation}\label{friedc}
H^2=\kappa\,\rho_{0}\, a(t)^{-(1+\omega)}.
\end{equation}
Equation (\ref{friedc}) contains Milne space $a(t)=\sqrt{\kappa\,\rho_{0}} \,t$ as a solution when $\omega=1$, i.e., for a radiation-dominated Universe (it is clear that a scale factor linear in time solves the second equation (\ref{ecsgr2}) as well when $\omega=1$). In order to be consistent with the previous notation, in the following we should call $a_{1}=\sqrt{\kappa\,\rho_{0}}$ (thus $a_{1}>0$ at this level). So, from a metric perspective, there is little more to say; Milne metric is a solution of the theory, and the diad $e^a_{\mu}=diag(1,a_{1} \,t)$ is not much more than a tool in order to get it. Furthermore, the value of $a_{1}>0$ is irrelevant; the coordinate change (\ref{coormil}) brings Milne metric into $ds^2=-(dx^{0})^2+(dx^{1})^2$ irrespective of the value of $a_{1}$. In other words, the geometry --the \emph{Milne wedge} $x^{0}>x^{1}$-- is not influenced by the constant $a_{1}$.

From a pure diad approach, in turn, the above conclusion is rather insufficient. The diad $e^a_{\mu}=diag(1,a_{1} \,t)$ is just one among many others describing the space under consideration. The whole set of admissible diads consistent with the metric $ds^2=-dt^2+a_{1}^2\,t^2dx^2$ (and hopefully, their physical meaning), should be an outcome of the theory as well. So, in order to comprehend the set $\mathcal{A}(e^a)$, we need to boost the (almost fortuitously found) \emph{seed} diad $e^a_{\mu}=diag(1,a_{1} \,t)$ and to ask for the invariance of $\mathbb{T}$. On the other hand, a characterization of $\mathcal{E}(Milne)$ will involve to solve the full equations for the entire diad field under the assumption of a linear scale factor. In the following paragraphs we will discuss these subjects. However, we should comment first on a subtle point. The relation $a_{1}=\sqrt{\kappa \,\rho_{0}}$ was a consequence of the Eqs. (\ref{ecsgr2}), which were obtained prior to the action of any Lorentz boost on the diad. So, when boosting the seed diad, as we are going to do in a moment, we need to contemplate the possibility of having another constant in front of the linear scale factor; in other words, the linear coefficient ${a}_{1}$ does not need to be $\sqrt{\kappa \,\rho_{0}}$ any longer. Its value (or rather, its form) will be dictated by the boosted equations.

In this way, if we locally boost Milne seed diad $e^a_{\mu}=diag(1,a_{1} \,t)$, we obtain
\begin{eqnarray}\label{diadboost}
e^{1}&=\cosh[\phi(t,x)]\,dt+a_{1} \,t\sinh[\phi(t,x)]\,dx,\notag\\
e^{2}&=\sinh[\phi(t,x)]\,dt+a_{1} \,t\cosh[\phi(t,x)]\,dx.
\end{eqnarray}
Here the \emph{booston} $\phi(t,x)$ appeared; its dependence on the spacetime coordinates should be kept in mind from now on. From (\ref{diadboost}) we compute the non-trivial torsion components
\begin{equation}\label{diadboosttor}
T^{t}_{\,\,\,\,t\,x}=a_{1}\,t\,\dot{\phi},\,\,\,\, T^{x}_{\,\,\,\,t\,x}=\frac{a_{1}-\phi'}{a_{1}\,t},
\end{equation}
where $\dot{\phi}=\partial\phi/\partial t$ and $\phi'=\partial\phi/\partial x$. Straightforwardly, we can obtain $\mathbb{T}$ as
\begin{equation}\label{eltmilne}
\mathbb{T}=\frac{2(a_{1}-\phi')^2}{a_{1}^2\,t^2}-2\dot{\phi}^2.
\end{equation}
This simple result permits us to fully characterize the remnant symmetries associated to Milne space. As a matter of fact, the value of $\mathbb{T}$ corresponding to the seed diad $e^a_{\mu}=diag(1,a_{1} \,t)$, this is $\mathbb{T}=2\,t^{-2}$, is unchanged provided the booston verifies
\begin{equation}\label{boostrem}
\phi(t,x)=\phi_{0}\,x + \phi_{t}\,\log(t)+\phi_{1},\,\,\,\,\,\,\phi_{t}^2=\frac{\phi_{0}}{a_{1}}\left(\frac{\phi_{0}}{a_{1}}-2\right),
\end{equation}
where $\phi_{0}$ and $\phi_{1}$ are integration constants. Then, apart from the full spacetime dependence, we have two cases of interest: (1) The constant (trivial) booston $\phi=\phi_{1}$, which is obtained by taking $\phi_{0}=0$. (2) The purely spatial, linear booston, $\phi(x)=2a_{1}\,x+\phi_{1}$, obtained by taking $\phi_t=0$, this is $\phi_{0}=2a_{1}$. Additionally, although no purely temporal booston arises as a remnant symmetry, it certainly does in the limit $a_{1}\rightarrow0$, $\phi_{0}\rightarrow0$, provided $\phi_{0}/a_{1}=\mathcal{O}(1)$ in that limit. However, we need to be aware that the limit $a_{1}\rightarrow0$ makes the ``metric sector" to blow up, except when $t\rightarrow \infty$, so this really makes sense only asymptotically in time.

\bigskip
Now, we will move on to what is more important, this is, the study of the set $\mathcal{E}(Milne)$, which involves the characterization of the full motion equations (\ref{eq: ecTT}) adapted to the boosted diad (\ref{diadboost}). Taking this into account, the equations result
\begin{eqnarray}
\phi''-a_{1}t\, \phi'\, \dot{\phi}&=&0,\label{ecsbaja11}\\
\ddot{\phi}+t^{-1}\dot{\phi}\,(1-a_{1}^{-1}\phi')&=&0,\label{ecsbaja21}\\
t^{-2}(1-b-a_{1}^{-2}\phi'^{\,2})+2\,a_{1}^{-1}t^{-1}\dot{\phi}'-\dot{\phi}^{2}&=&0,\label{ecsbaja31}
\end{eqnarray}
where $b=\kappa \,\rho_{0}\,a_{1}^{-2}$. This constant is thus representative of the matter content through the energy density. Of course, the constant booston trivially solves the system (\ref{ecsbaja11})-(\ref{ecsbaja31}) with $b=1$ (this is $a_{1}=\sqrt{\kappa \,\rho_{0}}$, as before). On the other hand, if $\phi=\phi(x)$ alone, then Eq. (\ref{ecsbaja21}) is immediately solved, while the other two give us
\begin{equation}\label{boosdepx}
\phi(x)=\phi_{0}\,x+\phi_{1},\,\,\,\,\,\,\phi_{0}^2=a_{1}^2-\kappa \,\rho_{0}=a_{1}^2(1-b),
\end{equation}
with constant $\phi_{1}$. We can inquire whether or not a purely temporal solution of the system (\ref{ecsbaja11})-(\ref{ecsbaja31}) actually exists. This seems to be the case, for if $\phi=\phi(t)$ alone, Eq. (\ref{ecsbaja11}) is solved at once, and the other two lead us to
\begin{equation}\label{boosdept}
\phi(t)=\tilde{\phi}_{0}\,\log(t) +\phi_{1},\,\,\,\,\,\,\tilde{\phi}_{0}^2=1-b.
\end{equation}
Finally, we can obtain yet another solution by changing to a new variable, namely
\begin{equation}\label{boosdepxt}
\phi(z)=\log(z) + z_{0},\,\,\,\,\,\,z=\frac{a_{1} x -\log(t)}{a_{1} x +\log(t)},\,\,\,\,\,b=1.
\end{equation}

We are now in position to discuss some similarities and differences present in both sets $\mathcal{A}(e^a)$ and $\mathcal{E}(Milne)$. Beyond the constant booston, which is an element belonging to both sets, we see that the general remnant booston (\ref{boostrem}) is solution of the equations (\ref{ecsbaja31}), only if $\phi_t=0$ and $\phi_{0}^2=a_{1}^2(1-b)$. This is just the linear booston $\phi=2a_{1}x+\phi_{1}$, but $a_{1}^2=-\kappa \,\rho_{0}/3$, which makes sense only if the energy density is negative. So, we conclude that linear, purely spatial symmetries belong to the two sets $\mathcal{A}(e^a)$ and $\mathcal{E}(Milne)$, but the coefficient of the linear term is different depending to what set we are looking at. Of course, what is ultimately important is the characterization of the set $\mathcal{E}(Milne)$, and the symmetries arising from the study of $\mathcal{A}(e^a)$ should be only a guide to know it. We see, thus, that $\mathcal{A}(e^a)$ really do its job partially by providing us a correct spatial sub-sector of $\mathcal{E}(Milne)$.

The relation between $\mathcal{A}(e^a)$ and $\mathcal{E}(Milne)$ in what concerns to the temporal sub-sector is even more tenuous. As mentioned in point (3) above, the temporal sub-sector exist only asymptotically at the level of the remnant symmetries. The identification $\phi_{t}^2\leftrightarrow \tilde{\phi}_{0}^2$ needed to put the two symmetries in contact (see Eq. (\ref{boosdept})) requires
\begin{equation}\label{iden}
 \frac{\phi_{0}}{a_{1}}\left(\frac{\phi_{0}}{a_{1}}-2\right)\leftrightarrow 1-  \frac{\kappa \,\rho_{0}}{a_{1}^2},
\end{equation}
as long as $a_{1}\rightarrow0$ and $\phi_{0}\rightarrow0$. This tells us that both symmetries exactly coincide when, besides these limits, we have also $\rho_{0}\rightarrow 0$ properly (vacuum case).

There is no way in which the spacetime booston (\ref{boosdepxt}) could have been anticipated from the remnant symmetries (\ref{boostrem}). It simply emerges as a pure symmetry belonging to $\mathcal{E}(Milne)$, but not to $\mathcal{A}(e^a)$. It is actually somewhat rare; it is valid only for $b=1$, this is, for $a_{1}^2=\kappa \,\rho_{0}$, just as the constant (global) booston. However, (\ref{boosdepxt}) is topologically disconnected from the global booston, and in particular, to the group identity. The global symmetry cannot be obtained from (\ref{boosdepxt}) by muting integration constants.

\subsection{Physical interpretation of $\mathcal{E}(Milne)$}

We can actually extract some physical meaning from the symmetries just found, at least in some cases. Here the coordinate changes associated to the three stages discussed in section \ref{sec2} will prove to be very valuable. In that opportunity, we were focused on how to understand the near-horizon geometry (i.e., the metric) on either side of the horizon. Now, instead, having obtained the boostons adapted to Milne metric (near horizon geometry viewed from the inside), we are interested in discovering how they look like from the exterior, Rindler-like description. It is important to take into account that no further Lorentz boost will be performed; we only need to know how the already obtained symmetries look from outside --and near to-- the horizon.

Just to be concise, let us consider the linear booston of Eq. (\ref{boosdepx}). We are behind the interpretation of the local symmetries, so let us fix $\phi_{1}=0$. The corresponding coordinate changes acting on the diad unfold in the following manner:

\bigskip

\emph{Stage one} (Eq. (\ref{change1})): from $(t,x)$ to $(\bar{T},x)$. The diad looks
\begin{eqnarray}\label{diadboostchange1}
e^{1}&= (2a_{1}\bar{T}-1)^{-1/2}\,\cosh[\phi_{0}\,x]\,d\bar{T}+(2a_{1}\bar{T}-1)^{1/2}\,\sinh[\phi_{0}\,x]\,dx,\notag\\
e^{2}&=(2a_{1}\bar{T}-1)^{-1/2}\,\sinh[\phi_{0}\,x]\,d\bar{T}+(2a_{1}\bar{T}-1)^{1/2}\,\cosh[\phi_{0}\,x]\,dx.
\end{eqnarray}

\emph{Stage two} (Eq. (\ref{metboostchange2})): from $(\bar{T},x)$ to $(T,\bar{x})$. Here we have
\begin{eqnarray}\label{diadboostchange2}
e^{1}&=(1-2a_{1}\bar{x})^{-1/2} \,\cosh[\phi_{0}\,T]\,d\bar{x}+(1-2a_{1}\bar{x})^{1/2}\,\sinh[\phi_{0}\,T]\,dT,\notag\\
e^{2}&=(1-2a_{1}\bar{x})^{-1/2} \,\sinh[\phi_{0}\,T]\,d\bar{x}+(1-2a_{1}\bar{x})^{1/2}\,\cosh[\phi_{0}\,T]\,dT.
\end{eqnarray}

\emph{Stage three} (Eq. (\ref{change3})): from $(T,\bar{x})$ to $(T,X)$. Finally, the diad acquires the form
\begin{eqnarray}\label{diadboostchange3}
e^{1}&=\cosh[\phi_{0}\,T]\,dX+(1-a_{1}X)\,\sinh[\phi_{0}\,T]\,dT,\notag\\
e^{2}&=\sinh[\phi_{0}\,T]\,dX+(1-a_{1}X)\,\cosh[\phi_{0}\,T]\,dT.
\end{eqnarray}
This last expression contains valuable information. Of course, the frame (\ref{diadboostchange3}) leads to the 2D Rindler line element of Eq. (\ref{change3}) for any value of $\phi_{0}=a_{1}\sqrt{1-b}$ (see Eq. (\ref{boosdepx})). However, when $\phi_{0}=a_{1}$ something peculiar happens, for in that case (\ref{diadboostchange3}) corresponds exactly to the diad carried by an uniformly accelerated observer with acceleration $a_{1}$ in Minkowski space. In other words, the frame
\begin{eqnarray}\label{diadboostchange31}
e^{1}&=\cosh[a_{1}T]\,dX+(1-a_{1}X)\,\sinh[a_{1}T]\,dT,\notag\\
e^{2}&=\sinh[a_{1}T]\,dX+(1-a_{1}X)\,\cosh[a_{1}T]\,dT,
\end{eqnarray}
is constructed from coordinates adapted to constant accelerated motion in flat space, see a detailed exposition in \cite{gravitation}. Remember that $a_{1}=1/4M$ is the surface gravity, so the bigger the black hole mass, the smaller the acceleration experienced near the horizon. Interesting enough is the fact that this interpretation occurs only when $b=0$, which corresponds to the vacuum case. This is possible because Milne space emerges as a solution of the equations (\ref{eq: ecTT}) also in the vacuum case, at the expense of having local extra degrees of freedom encoded in the diad. In other words, a sector of $\mathcal{E}(Milne)$ (linear booston) generates not only the Rindler near-horizon geometry experienced by a radial observer in the Schwarzschild black hole, but it also can be consistently interpreted as an observer in \emph{constant} accelerated motion in pure vacuum. This happens as well for the temporal booston of Eq. (\ref{boosdept}), but the acceleration has to be fixed to an unitary value in that case, see Ref. \cite{Andronikos}.

\section{Far from the horizon: a quick look to Minkowski space}

The simplicity of 2D gravity allows to go further and to face yet another fundamental question. What is Minkowski space according to these models? This question is not only important for determining a correct vacuum, but also to figure out the links between local and global aspects within these frameworks. This very question was raised before in the context of 4D-$f(T)$ gravity, see some developments in Ref. \cite{Golo2}.

The full set of diads adapted to 2D Minkowski space is basically the result of a local Lorentz boost applied to the \emph{Euclidean} diad $e^a_{\,\,\mu}=\delta^a_{\,\,\nu}$, this is  $e^b_{\,\,\nu}=\Delta^b_{\,\,a}\, \delta^a_{\,\,\nu}$. Then we have
\begin{eqnarray}\label{diadboostmin}
e^{1}&=\cosh[\phi(t,x)]\,dt+\sinh[\phi(t,x)]\,dx,\notag\\
e^{2}&=\sinh[\phi(t,x)]\,dt+\cosh[\phi(t,x)]\,dx,
\end{eqnarray}
which plugged into (\ref{eq: ecTT}) give us
\begin{equation}
\phi''-\phi'\, \dot{\phi}=0,\,\,\,\,\,\,\,\,\ddot{\phi}-\phi'\,\dot{\phi}=0,\,\,\,\,\,\,\,\,\dot{\phi}^2-2\dot{\phi}'+\phi'^{\,2}=0.\label{ecsbaja}
\end{equation}
It is clear that no purely spatial or temporal solutions exist. In turn, the first two equations imply $\square^2\phi=0$, which is solved in outgoing and ingoing null coordinates $V=t-x$ and  $U=t+x$ by $\phi=\phi_{V}-\phi_{U}$, where $\phi_{U}\equiv\phi_{U}(U)$ and $\phi_{V}\equiv\phi_{V}(V)$ are arbitrary (twice differentiable) functions of the null coordinates. If we then replace $\phi$ in the last equation (\ref{ecsbaja}), we find
\begin{equation}
\phi_{U}=\log(u_{0}\,U+u_{1}),\,\,\, \phi_{V}=\log(v_{0}\,V+v_{1}),
\label{ondassol}
\end{equation}
which automatically solves the first two for arbitrary constants $u_{i},v_{i}$, $i=0,1$. These are the boostons adapted to Minkowski space, i.e., the set $\mathcal{E}(Minkowski)$. The results are clearer if we write down the diad (\ref{diadboostmin}) in null coordinates. If we use $Cosh[Log(y)]=(1+y^2)/2y$ and $Sinh[Log(y)]=(-1+y^2)/2y$, we can split the diad into two independent propagating ``modes", namely
\begin{equation}\label{modoout}
e^{1}_{out}=\frac{dU+dV}{2(v_{0}\,V+v_{1})},\,\,\,\,\,e^{2}_{out}=\frac{dU-dV}{2(v_{0}\,V+v_{1})}\,,
\end{equation}
\begin{equation}\label{modoin}
e^{1}_{in}=\frac{dU+dV}{2(u_{0}\,U+u_{1})},\,\,\,\,\,e^{2}_{in}=\frac{dU-dV}{2(u_{0}\,U+u_{1})}\,.
\end{equation}
According to this model, Minkowski space is then characterized by the propagating modes (\ref{modoout}) and (\ref{modoin}), which \emph{generate} the metric $ds^2=-dUdV$. The entire diad is parametrized by the constants $v_{0}$ and $u_{0}$; global, non propagating diads require $v_{0}=u_{0}=0$. However, an important point should be kept in mind: the solutions (\ref{modoout}) and (\ref{modoin}) are not valid on the past and future null cones defined by $V=-v_{1}/v_{0}$ and $U=-u_{1}/u_{0}$ provided $v_{0}\neq0$ and $u_{0}\neq0$. This is not a problem of the coordinates, it is in fact an intrinsic property of the solutions. In this sense, a global well defined parallelization, this is, a global, smooth fields of diads is only obtained when $v_{0}$ and $u_{0}$ are both null (global booston). Hence, it is obvious that the set $\mathcal{E}(Minkowski)$ contains more than just geometry, because it carries information concerning global aspects.

On the other hand we can easily compute $\mathbb{T}$ from (\ref{diadboostmin}), obtaining the limpid result
\begin{equation}
\mathbb{T}=-2 \,\eta^{\mu\nu}\phi_{,\mu}\,\phi_{,\nu}.
\label{limpidl}
\end{equation}
Due to the fact that $\mathbb{T}=0$ for the Euclidean diad, the remnant symmetries require $\eta^{\mu\nu}\phi_{,\mu}\,\phi_{,\nu}=0$, which is satisfied for arbitrary functions $f(U)$ and $g(V)$ of the null coordinates separately, but not for a combination of the two. In this way, the boostons (\ref{ondassol}) are in $\mathcal{A}(\delta^a_{\,\,\nu})$, but $\phi=\phi_{V}-\phi_{U}$ is not a booston belonging to $\mathcal{A}(\delta^a_{\,\,\nu})$. Apparently, this is the closest we can be of establishing a link between $\mathcal{A}(\delta^a_{\,\,\nu})$ and $\mathcal{E}(Minkowski)$.

\section{Final comments}

Simple models for 2D-torsional gravity have shown several enlightening results in what concerns to the understanding of the local symmetries arising in the study of their 4D counterparts. Once we have acknowledged that the near horizon, Schwarzschild geometry adapted to radial motion is inextricably related to 2D Milne model, the theory described by the action (\ref{accionlow}) and the equations (\ref{eq: ecTT})  was fully characterized under those circumstances. To \emph{fully characterized} Milne space in this context means that one is capable to obtain from the motion equations, all the diads compatible with Milne metric. This set of diads, which we entitled $\mathcal{E}(Milne)$, is given by (\ref{diadboost}) with the boostons $\phi(t,x)$ of Eqs. (\ref{boosdepx}), (\ref{boosdept}) and (\ref{boosdepxt}). Furthermore, the simplicity of the 2D action enables to inquire on which LLT leave invariant the value of $\mathbb{T}$ corresponding to the Milne seed diad, namely, the remnant group associated to Milne space, $\mathcal{A}(Milne)$.

An important lesson coming from the characterization of these two sets is the fact that they are essentially different in general, even though they may share some elements. This is an important conclusion because normally it is much easier to obtain $\mathcal{A}(e^a)$ than $\mathcal{E}(\textbf{g})$, but we do know now that to assume elements belonging to the latter by having (even partial) information about the former, could be definitely wrong. It is worth of noticing that the general discrepancy between the two sets comes not from freezing degrees of freedom at the level of the action --i.e., it is not due to \emph{symmetric criticality} violations \cite{Palais}-- but as a consequence of the manifestly non local Lorentz invariant character of the action. Some connections between the two sets remain, though. This is evident for the null $\mathbb{T}$ representative of the Euclidean diad associated to Minkowski space. Null modes are remnant symmetries which, in turn, are further constrained by the motion equations, as Eq. (\ref{ondassol}) shows.

At the end, what is important is the set $\mathcal{E}(\textbf{g})$, and its interpretation beyond the metric point of view. By means of the Milne-Rindler ``bridge", which links the near-horizon description on either side of the Schwarzschild horizon, we were able to extract a physical interpretation of the preferred diads arising as a consequence of the breaking of the local Lorentz invariance. Some of the local symmetries, as  (\ref{boosdepx}) and (\ref{boosdept}), represent uniformly accelerated observers whose acceleration is given by the surface gravity of the Schwarzschild black hole. In this way, the action (\ref{accionlow}) contains a sector controlling the local geometry (this sector is governed by the contracted equations (\ref{eq: ecTTcon}) conducing to the JT equation (\ref{JT})), and another one dictating the full orientation of the diad. In a way, this last sector tells us how a free falling observer should perceive the local geometry. Both aspects, the local geometry and the state of motion, are encoded in the diad field. In this sense, the set $\mathcal{E}(\textbf{g})$ contains more than geometry, because global aspects --like the presence of a Rindler horizon-- are also contained in $\mathcal{E}(\textbf{g})$. In GR, those global aspects are put by hand, so to speak; they appear as intricate coordinates changes which restrict the range of certain charts, as the Rindler chart, for instance. Perhaps it would be helpful in the future to think more frequently on this global-local marriage when dealing with tetrad-based, non-local Lorentz invariant theories as $f(T)$ gravity.

\bigskip

\textbf{\emph{Acknowledgements}}. I want to express my gratitude to the organizers of the Metric-Affine Gravity conference held at Tartu last year, for the nice forum they put together. I also want to offer them my apologies for not being there more than virtually. The author is member of \emph{Carrera del Investigador Cient\'{i}fico} (CONICET), and his work is supported by CONICET and Instituto Balseiro (UNCUYO).

\end{document}